\documentclass[aps,prb,twocolumn,showpacs,floatfix]{revtex4}

\usepackage{graphicx}

\begin{document}

\title{Magnetic susceptibility of the two-dimensional Hubbard model\\ 
using a power series for the hopping constant}

\author{Alexei Sherman}
 \affiliation{Institute of Physics, University of Tartu, Riia 142,
 51014 Tartu, Estonia}
\author{Michael Schreiber}
 \affiliation{Institut f\"ur Physik, Technische Universit\"at, 09107
 Chemnitz, Germany}

\date{\today}

\begin{abstract}
The magnetic susceptibility of the two-dimensional repulsive Hubbard
model with nearest-neighbor hopping is investigated using the diagram
technique developed for the case of strong correlations. In this
technique a power series in the hopping constant is used. At
half-filling the calculated zero-frequency susceptibility and the
square of the site spin reproduce adequately results of Monte Carlo
simulations. Also in agreement with numerical simulations no evidence
of ferromagnetic correlations was found in the considered range of
electron concentrations $0.8\alt\bar{n}\alt 1.2$ for the repulsion
parameters $8|t|\leq U\leq 16|t|$. However, for larger $U/|t|$ and
$|1-\bar{n}|\approx 0.2$ the nearest neighbor correlations become
ferromagnetic. For $\bar{n}\alt 0.94$ and $\bar{n}\agt 1.06$ the
imaginary part of the real-frequency susceptibility becomes
incommensurate for small frequencies. The incommensurability parameter
grows with departure from half-filling and decreases with increasing
the frequency. This behavior of the susceptibility can explain the
observed low-frequency incommensurate response observed in normal-state
lanthanum cuprates.
\end{abstract}

\pacs{71.10.Fd, 71.27.+a, 75.40.Gb}

\maketitle

\section{Introduction}
The Hubbard model\cite{Hubbard} is thought to be appropriate to
describe the main features of electron correlations in narrow energy
bands, leading to collective effects such as magnetism and
metal-insulator transition. It has been often used to describe real
materials exhibiting these phenomena (see, e.g.,
Refs.~\onlinecite{Izyumov}, \onlinecite{Ovchinnikov}, and references
therein).

In more than one dimension, the model is not exactly solvable and a
variety of numerical and analytical approximate methods was used for
its study. Among others there are Monte Carlo
simulations,\cite{Hirsch,Moreo} different cluster methods,\cite{Maier}
the composite operator formalism,\cite{Mancini} the generating
functional approach,\cite{Izyumov05} Green's function decoupling
schemes,\cite{Irkhin} and variational approaches.\cite{Seibold} Along
with these methods various versions of the diagram technique
\cite{Izyumov,Ovchinnikov,Zaitsev,Vladimir,Pairault,Sherman06} have
been used for the investigation of the model. In the case of strong
electron correlations when the ratio of the hopping constant $t$ to the
on-site repulsion $U$ is a small parameter the use of the diagram
technique based on the series expansion in this parameter is quite
reasonable.

In the present work we use the diagram technique of
Refs.~\onlinecite{Vladimir} and \onlinecite{Sherman06} for
investigating the magnetic susceptibility of the one-band
two-dimensional repulsive Hubbard model with nearest-neighbor hopping
in the case of strong electron correlations. In this version of the
diagram technique terms of the power expansion are expressed through
cumulants of creation and annihilation electron operators. The
considered model possesses the electron-hole symmetry and results
obtained for electron concentrations $\bar{n}<1$ are replicated for
$\bar{n}>1$. Therefore in the following discussion we shall restrict
our consideration to the former region of concentrations.

We found that at half-filling the calculated temperature dependence of
the zero-frequency susceptibility reproduces adequately key features of
results of Monte Carlo simulations.\cite{Hirsch} The uniform
susceptibility tends to a finite value for vanishing temperature. The
staggered susceptibility diverges with decreasing temperature which
signals the establishment of the long-range antiferromagnetic order.
The transition temperature $T_0$ is finite which indicates the
violation of the Mermin-Wagner theorem.\cite{Mermin} However, the
transition temperature is always lower than the analogous temperature
in the random phase approximation (RPA). Besides, the transition
temperature decreases with decreasing the ratio $|t|/U$ of the hopping
constant and the on-site repulsion, i.e.\ the violation of the
Mermin-Wagner theorem becomes less pronounced on enforcing the
condition for which the approximation was developed. For small ratios
$|t|/U$ the calculated square of the site spin differs by less than
10\% from the data of Monte Carlo simulations. Also in agreement with
Monte Carlo results we found no evidence of ferromagnetic correlations
in the considered range of electron concentrations $0.8\alt\bar{n}\alt
1.2$ for the repulsion parameters $8|t|\leq U\leq 16|t|$. However, for
larger $U/|t|$ and $|1-\bar{n}|\approx 0.2$ the nearest neighbor
correlations become ferromagnetic. In the case $U=8|t|$ for
$0.94\alt\bar{n}\alt 1.06$ the zero-frequency susceptibility and the
imaginary part of the susceptibility for low real frequencies are
peaked at the antiferromagnetic wave vector $(\pi,\pi)$. For smaller
and larger concentrations these susceptibilities become incommensurate
-- momenta of their maxima deviate from $(\pi,\pi)$ -- and the
incommensurability parameter, i.e.\ the distance between $(\pi,\pi)$
and the wave vector of the susceptibility maximum, grows with departure
from half-filling. With increasing the frequency the incommensurability
parameter decreases and finally vanishes. This behavior of the strongly
correlated system resembles the incommensurate magnetic response
observed in the normal-state lanthanum cuprates\cite{Yamada} and can be
used for its explanation.

Main formulas used in the calculations are given in the following
section. The discussion of the obtained results and their comparison
with data of Monte Carlo simulations are carried out in Sec.~III\@.
Concluding remarks are presented in Sec.~IV. A relation between the
longitudinal and transversal spin Green's function is checked in the
Appendix.
\begin{figure}
\centerline{\includegraphics[width=6.5cm]{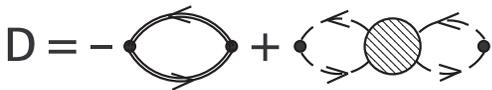}} \caption{The
diagram equation for $D({\bf k},i\omega_\nu)$.} \label{Fig_i}
\end{figure}

\section{Main formulas}
The Hubbard model is described by the Hamiltonian
\begin{equation}\label{Hamiltonian}
H=\sum_{\bf ll'\sigma}t_{\bf ll'}a^\dagger_{\bf l\sigma}a_{\bf
 l'\sigma}+\frac{U}{2}\sum_{\bf l\sigma}n_{\bf l\sigma}n_{\bf
 l,-\sigma},
\end{equation}
where $a^\dagger_{\bf l\sigma}$ and $a_{\bf l\sigma}$ are the electron
creation and annihilation operators, ${\bf l}$ labels sites of the
square plane lattice, $\sigma=\pm 1$ is the spin projection, $t_{\bf
ll'}$ and $U$ are hopping and on-site repulsion constants, and $n_{\bf
l\sigma}=a^\dagger_{\bf l\sigma}a_{\bf l\sigma}$. Below we consider the
case where only the constant $t$ for hopping between nearest neighbor
sites is nonzero.

In the case of strong correlations, $U\gg|t|$, for calculating Green's
functions it is reasonable to use the expansion in powers of the
hopping constant. In the diagram technique of
Refs.~\onlinecite{Vladimir} and \onlinecite{Sherman06} this expansion
is expressed in terms of site cumulants of electron creation and
annihilation operators. We use this technique for calculating the spin
Green's function
\begin{equation}\label{Green}
D({\bf l'\tau',l\tau})=\langle{\cal T}s^\sigma_{\bf l'}(\tau')
s^{-\sigma}_{\bf l}(\tau)\rangle,
\end{equation}
where $s^\sigma_{\bf l}=a^\dagger_{\bf l\sigma}a_{\bf l,-\sigma}$ is
the spin operator, the angular brackets denote the statistical
averaging with the Hamiltonian
$${\cal H}=H-\mu\sum_{\bf l\sigma}n_{\bf
l\sigma},$$ $\mu$ is the chemical potential, ${\cal T}$ is the
time-ordering operator which arranges other operators from right to
left in ascending order of times $\tau$, and
$$a_{\bf
l\sigma}(\tau)=\exp({\cal H}\tau)a_{\bf l\sigma}\exp(-{\cal H}\tau).$$

\begin{figure}
  \centerline{\includegraphics[width=6.5cm]{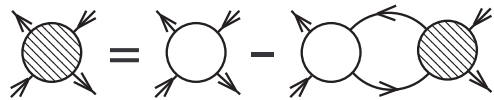}}
  \caption{The Bethe-Salpeter equation for the sum of all four-leg
  diagrams.}\label{Fig_ii}
\end{figure}

The structure elements of the used diagram technique are site cumulants
and hopping constants which connect the cumulants.
\cite{Vladimir,Sherman06} In diagrams, we denote the hopping constants
by single directed lines. Using the diagram technique it can be shown
that Green's function (\ref{Green}) satisfies the diagram equation
plotted in Fig.~\ref{Fig_i}. In this diagram, after the Fourier
transformation over the space and time variables the dual line
indicates the full electron Green's function
$$
G({\bf k},n)=-\frac{1}{2}\int_{-\beta}^\beta e^{i\omega_n\tau}
\big\langle{\cal T}a_{\bf k}(\tau)a^\dagger_{\bf k}\big\rangle\, d\tau,
$$
where ${\bf k}$ is the wave vector, the integer $n$ stands for the
fermion Matsubara frequency $\omega_n=(2n+1)\pi T$ with the temperature
$T$, and $\beta=T^{-1}$. The shaded circle in Fig.~\ref{Fig_i} is the
sum of all four-leg diagrams, i.e. such diagrams in which starting from
any leg one can reach any other leg moving along the hopping lines and
cumulants. These diagrams can be separated into reducible and
irreducible diagrams. In contrast to the latter, the reducible diagrams
can be divided into two disconnected parts by cutting two hopping
lines. The sum of all four-leg diagrams satisfies the Bethe-Salpeter
equation shown in Fig.~\ref{Fig_ii}. Here the open circle indicates the
sum of all irreducible four-leg diagrams. The hopping lines between the
open and shaded circles are already renormalized here by the inclusion
of all possible irreducible two-leg diagrams into these lines. These
irreducible two-leg diagrams cannot be divided into two disconnected
parts by cutting one hopping line.\cite{Sherman06} As a consequence,
the hopping line in Fig.~\ref{Fig_ii} is described by the equation
\begin{equation}\label{hopping}
\Theta({\bf k},n)=t_{\bf k}+t_{\bf k}^2G({\bf k},n),
\end{equation}
where in the considered model with nearest-neighbor hopping we have
$t_{\bf k}=2t[\cos(k_x)+\cos(k_y)]$. The irreducible two-leg diagrams
can also be inserted in the external lines of the four-leg
diagrams in Fig.~\ref{Fig_i}. To mark this renormalization we use
dashed lines in that figure. Each of these lines introduces the
multiplier $\Pi({\bf k},n)=\Theta({\bf k},n)/t_{\bf k}$ in the second
term on the right-hand side of the equation in Fig.~\ref{Fig_i}.
Without the renormalization this multiplier reduces to unity.

As a result, the equations depicted in Figs.~\ref{Fig_i} and
\ref{Fig_ii} read
\begin{eqnarray}
D(p)&=&-N^{-1}T\sum_{p_1}G(p_1)G(p+p_1)\nonumber\\
&+&N^{-2}T^2\sum_{p_1p_2}\Pi(p_1)\Pi(p_2)\Pi(p+p_1)\Pi(p+p_2)
 \nonumber\\
&\times&\Gamma(p_1,p+p_1,p+p_2,p_2), \label{GF}
\end{eqnarray}
\begin{eqnarray}
&&\Gamma(p_1,p+p_1,p+p_2,p_2)=\gamma(p_1,p+p_1,p+p_2,p_2)\nonumber\\
&&\quad-N^{-1}T\sum_{p_3}\gamma(p_1,p+p_1,p+p_3,p_3)\Theta(p_3)
 \Theta(p+p_3)\nonumber\\
&&\quad\times\Gamma(p_3,p+p_3,p+p_2,p_2). \label{Bethe}
\end{eqnarray}
Here the combined indices $p=({\bf k},i\omega_\nu)$ and $p_j=({\bf
k}_j,i\omega_{n_j})$ were introduced, $\omega_\nu=2\nu\pi T$ is the
boson Matsubara frequency, $\Gamma(p_1,p+p_1,p+p_2,p_2)$ is the sum of
all four-leg diagrams, $\gamma(p_1,p+p_1,p+p_2,p_2)$ is its irreducible
subset, and $N$ is the number of sites.

In the following consideration we simplify the general
equations~(\ref{GF}) and (\ref{Bethe}) by neglecting the irreducible
two-leg diagrams in the external and internal lines of the four-leg
diagrams and by using the lowest-order irreducible four-leg diagram
instead of $\gamma(p_1,p+p_1,p+p_2,p_2)$. This four-leg diagram is
described by the second-order cumulant
\begin{eqnarray}
K_2(\tau',\tau,\tau'_1,\tau_1)&=&\langle{\cal T}\bar{a}_\sigma(\tau')
 a_{-\sigma}(\tau)
 \bar{a}_{-\sigma}(\tau'_1)a_{\sigma}(\tau_1)\rangle_0\nonumber\\
&+&K_1(\tau',\tau_1)K_1(\tau'_1,\tau) \label{cumulant},
\end{eqnarray}
where the subscript ``0'' of the angular bracket indicates that the
averaging and time dependencies of the operators are determined by the
site Hamiltonian
\begin{eqnarray*}
&&H_{\bf l}=\sum_\sigma[(U/2)n_{\bf l\sigma}n_{\bf l,-\sigma}-\mu
n_{\bf l\sigma}],\\
&&\bar{a}_{\bf l\sigma}(\tau)= \exp(H_{\bf l}\tau)a^\dagger_{\bf
l\sigma}\exp(-H_{\bf l}\tau),
\end{eqnarray*}
and the first-order cumulant $$K_1(\tau',\tau)=\langle{\cal
T}\bar{a}_\sigma (\tau')a_\sigma(\tau)\rangle_0.$$
All operators in the
cumulants belong to the same lattice site. Due to the translational
symmetry of the problem the cumulants do not depend on the site index
which is therefore omitted in the above equations. The expression for
$K_2$ reads
\begin{widetext}
\begin{eqnarray}
&&K_2(n_1,n_1+\nu,n_2+\nu,n_2)=Z^{-1}\Big\{\beta\big[\delta_{\nu,0}
 e^{-E_1\beta}+Z^{-1}\delta_{n_1,n_2}\big(e^{-2E_1\beta}-
 e^{-(E_0+E_2)\beta}\big)\big]F(n_1+\nu)F(n_2)\nonumber\\
&&\quad+e^{-E_0\beta}Ug_{01}(n_1+\nu)g_{01}(n_2)g_{02}(n_1+n_2+\nu)
 \big[g_{01}(n_2+\nu)+g_{01}(n_1)\big]\nonumber\\
&&\quad+e^{-E_2\beta}Ug_{12}(n_1+\nu)g_{12}(n_2)g_{02}(n_1+n_2+\nu)
 \big[g_{12}(n_2+\nu)+g_{12}(n_1)\big]\nonumber\\
&&\quad-e^{-E_1\beta}\Big[F(n_1+\nu)g_{01}(n_2)g_{01}(n_2+\nu)+
 F(n_2)g_{01}(n_1+\nu)g_{01}(n_1)\nonumber\\
&&\quad+F(n_2)g_{12}(n_2+\nu)\big[g_{12}(n_1+\nu)-g_{01}(n_1)\big]+
 F(n_1+\nu)g_{12}(n_1)\big[g_{12}(n_2)-g_{01}(n_2+\nu)\big]\Big]\Big\},
\label{K2}
\end{eqnarray}
where $E_0=0$, $E_1=-\mu$, and $E_2=U-2\mu$ are the eigenenergies of
the site Hamiltonian $H_{\bf l}$, $Z=e^{-E_0\beta}+
2e^{-E_1\beta}+e^{-E_2\beta}$ is the site partition function,
$g_{ij}(n)=(i\omega_n+E_i- E_j)^{-1}$, and $F(n)=g_{01}(n)-g_{12}(n)$.

It is worth noting that the used approximation retains the relation
\begin{equation}\label{rotinvar}
D({\bf l'\tau',l\tau})=2D_z({\bf l'\tau',l\tau}),
\end{equation}
where
\begin{equation}\label{Dz}
D_z({\bf l'\tau',l\tau})=\langle{\cal T}s^z_{\bf l'}(\tau')s^z_{\bf
l}(\tau) \rangle
\end{equation}
and $s^z_{\bf l}=\frac{1}{2}\sum_\sigma\sigma a^\dagger_{\bf l\sigma}
a_{\bf l\sigma}$ is the $z$ component of spin.
Relation~(\ref{rotinvar}) follows from the invariance of
Hamiltonian~(\ref{Hamiltonian}) with respect to rotations of the spin
quantization axis.\cite{Fradkin} The proof of Eq.~(\ref{rotinvar}) is
given in the Appendix.

Equation~(\ref{K2}) can be significantly simplified for the case of
principal interest $U\gg T$. In this case, if $\mu$ satisfies the
condition
\begin{equation}\label{condition}
\varepsilon<\mu<U-\varepsilon,
\end{equation}
where $\varepsilon\gg T$, the exponent $e^{-\beta E_1}$ is much larger
than $e^{-\beta E_0}$ and $e^{-\beta E_2}$. Therefore terms with
$e^{-\beta E_0}$ and $e^{-\beta E_2}$ can be omitted in Eq.~(\ref{K2})
which gives
\begin{eqnarray}
&&K_2(n_1,n_1+\nu,n_2+\nu,n_2)=\frac{1}{2}\Big\{\beta\Big(\delta_{\nu,0}
 +\frac{1}{2}\delta_{n_1,n_2}\Big)F(n_1+\nu)F(n_2)
 \nonumber\\
&&\quad-F(n_1+\nu)g_{01}(n_2)g_{01}(n_2+\nu)
 -F(n_2)g_{01}(n_1+\nu)g_{01}(n_1)\nonumber\\
&&\quad-F(n_2)g_{12}(n_2+\nu)\big[g_{12}(n_1+\nu)-g_{01}(n_1)\big]
 -F(n_1+\nu)g_{12}(n_1)\big[g_{12}(n_2)-g_{01}(n_2+\nu)\big]\Big\}.
\label{K2s}
\end{eqnarray}

From Eq.~(\ref{Bethe}) with the kernel~(\ref{K2s}) it can be seen that
$\Gamma$ does not depend on momenta ${\bf k}_1$ and ${\bf k}_2$. Since
we neglected irreducible diagrams in the external lines, $\Pi(p)=1$ and
in the second term on the right-hand side of Eq.~(\ref{GF}) the
summations over ${\bf k}_1$, ${\bf k}_2$, and $n_2$ can be carried out
instantly. The resulting equation for $\Gamma'_{\bf
k}(\nu,n)=T\sum_{n'}\Gamma_{\bf k}(n,n+\nu,n'+\nu,n')$ reads
\begin{eqnarray}
\Gamma'_{\bf k}(\nu,n)&=&\frac{1}{2}f_{\bf k}(\nu,n)\big\{2K'_2(\nu,n)+
 \big[a_2(-\nu,\nu+n)-a_1(\nu+n)\beta\delta_{\nu,0}\big]tt_{\bf k}
 y_1({\bf k}\nu)+a_1(\nu+n)tt_{\bf k}y_2({\bf k}\nu)\nonumber\\
&+&a_4(-\nu,\nu+n)tt_{\bf k}y_3({\bf k}\nu)+a_3(-\nu,\nu+n)tt_{\bf
k}y_4({\bf k}\nu)\big\} \label{Gamma},
\end{eqnarray}
where
\begin{eqnarray}
K'_2(\nu,n)&=&T\sum_{n'}K_2(n,n+\nu,n'+\nu,n')\nonumber\\
&=&\frac{1}{2}\bigg\{\bigg[\beta\delta_{\nu,0}+\frac{1}{2}
 a_1(n)\bigg]a_1(n+\nu)-a_2(-\nu,n+\nu)
 +\frac{1}{U-i\omega_\nu} a_4(-\nu,n+\nu)+a_3(-\nu,n+\nu)\bigg\},
\label{Kprime}
\end{eqnarray}
\begin{eqnarray}
&&f_{\bf k}(\nu,n)=\bigg[1+\frac{1}{4}F(n)F(\nu+n)tt_{\bf k}
 \bigg]^{-1}, \quad
 y_i({\bf k}\nu)=T\sum_n a_i(\nu,n)\Gamma'_{\bf k}(\nu,n),
 \nonumber\\[-3ex]
&&\label{terms}\\
&&a_1(n)=F(n),\quad
 a_2(\nu,n)=g_{01}(n)g_{01}(\nu+n),\quad
 a_3(\nu,n)=F(n)g_{12}(\nu+n),\quad
 a_4(\nu,n)=g_{12}(n)-g_{01}(\nu+n).\nonumber
\end{eqnarray}

Multiplying Eq.~(\ref{Gamma}) by $a_i(\nu,n)$ and summing over $n$ we
obtain a system of four linear algebraic equations for $y_i$,
\begin{equation}\label{system}
y_i=b_i+(c_{i2}-c_{i1}\beta\delta_{\nu,0})y_1+c_{i1}y_2+c_{i4}y_3+
c_{i3}y_4,
\end{equation}
where
$$b_i=T\sum_n a_i(\nu,n)K'_2(\nu,n)f_{\bf k}(\nu,n),\quad
 c_{ij}=tt_{\bf k}\frac{T}{2}\sum_n a_i(\nu,n)a_j(-\nu,\nu+n)f_{\bf
 k}(\nu,n).$$
System~(\ref{system}) can easily be solved. Thus, in the used
approximation the Bethe-Salpeter equation~(\ref{Bethe}) can be solved
exactly. In notations~(\ref{terms}) the second term on the right-hand
side of Eq.~(\ref{GF}) can be rewritten as
\begin{eqnarray}
\bigg(\frac{T}{N}\bigg)^2\sum_{p_1p_2}\Gamma&=&\frac{T}{2}\bigg\{\big[
 \beta\delta_{\nu,0}(1-tt_{\bf k}y_1)+tt_{\bf k}y_2\big]
 \sum_n f_{\bf k}(\nu,n)a_1(n+\nu)
 +\frac{1}{2}\sum_n f_{\bf k}(\nu,n)a_1(n)a_1(n+\nu)
 \nonumber\\
&-&(1-tt_{\bf k}y_1)\sum_n f_{\bf k}(\nu,n)a_2(-\nu,n+\nu)
 +\bigg(tt_{\bf k}y_3+\frac{1}{U-i\omega_\nu}\bigg)\sum_n f_{\bf
 k}(\nu,n)a_4(-\nu,n+\nu)\nonumber\\
&+&(1+tt_{\bf k}y_4)\sum_n f_{\bf
 k}(\nu,n)a_3(-\nu,n+\nu)\bigg\}.\label{second}
\end{eqnarray}
\end{widetext}

In subsequent calculations we shall use the Hubbard-I approximation
\cite{Hubbard} for the electron Green's function in the first term on
the right-hand side of Eq.~(\ref{GF}). In the used diagram technique
this approximation is obtained if in the Larkin equation the sum of all
irreducible two-leg diagrams is substituted by the first-order
cumulant.\cite{Vladimir,Sherman06} Provided that
condition~(\ref{condition}) is fulfilled the electron Green's function
in the Hubbard-I approximation reads
\begin{equation}\label{HubbardI}
G({\bf k}n)=\frac{i\omega_n+\mu-U/2}{(i\omega_n+\mu)(i\omega_n+\mu-U)
 -t_{\bf k}(i\omega_n+\mu-U/2)}.
\end{equation}

\section{Magnetic susceptibility}
From the Lehmann representation\cite{Mahan} it can be shown that
$D_z({\bf k}\nu)$ has to be real, nonnegative,
\begin{equation}\label{DRI}
D_z({\bf k}\nu)\ge 0
\end{equation}
and symmetric, $ D_z({\bf k}\nu)=D_z({\bf k},-\nu)$. In view of
Eq.~(\ref{rotinvar}) analogous relations are fulfilled for $D({\bf
k}\nu)$. However, we found that condition~(\ref{DRI}) is violated for
$\nu=0$ and some momentum if the temperature drops below some critical
value $T_0$ which depends on the ratio $|t|/U$ and on $\mu$. As the
temperature $T_0$ is approached from above, $D({\bf k},0)$ tends to
infinity which leads to the establishment of long-range spin
correlations. Therefore, like in the RPA,\cite{Mahan,Izyumov90} we
interpret this behavior of Green's function as a transition to a
long-range order. Near half-filling the highest temperature $T_0$
occurs for the antiferromagnetic momentum $(\pi,\pi)$. Thus, near
half-filling the system exhibits transition to the state with the
long-range antiferromagnetic order.

In our calculations $T_0$ is finite. Since we consider the
two-dimensional model and the broken symmetry is continuous, this
result is in contradiction to the Mermin-Wagner theorem\cite{Mermin}
and shows that the used approximation somewhat overestimates the effect
of the interaction. However, it is worth noting that the value of $T_0$
decreases with decreasing the ratio $|t|/U$, i.e. the violation of the
Mermin-Wagner theorem becomes less pronounced on enforcing the
condition for which the approximation was developed. Notice that other
approximate methods, including RPA\cite{Hirsch} and cluster methods,
\cite{Maier} lead also to the violation of the Mermin-Wagner theorem.
In the following calculations we consider only the region $T>T_0$.

It was also found that for $\nu\ne 0$ condition~(\ref{DRI}) is violated
in a small area of the Brillouin zone near the $\Gamma$ point. Green's
function is small for such momenta and small negative values of $D({\bf
k}\nu)$ here are a consequence of the used approximations. It is worth
noting that the renormalization of internal and external hopping lines
should improve the behavior of $D({\bf k}\nu)$ in this region.

\begin{figure}[t]
\centerline{\includegraphics[width=7.5cm]{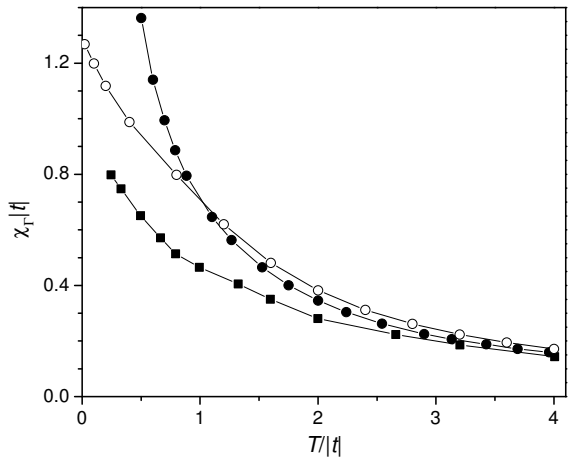}} \caption{The
zero-frequency magnetic susceptibility at ${\bf k}=0$ vs.\ temperature
at half-filling and $t=-U/4$. Filled squares, filled and open circles
are results of the Monte Carlo simulations, \protect\cite{Hirsch}
random phase approximation, and our calculations,
respectively.}\label{Fig_iii}
\end{figure}
To check the used approximation we shall compare our calculated results
with data of Monte Carlo simulations\cite{Hirsch} on the temperature
dependence of the zero-frequency susceptibility at half-filling and on
the square of the site spin $\langle{\bf S}^2\rangle$. In the usual
definition\cite{Mahan} the susceptibility $\chi({\bf k}\nu)$ differs
from $D({\bf k}\nu)$ only in a constant multiplier. For convenience in
comparison with results of Ref.~\onlinecite{Hirsch} in this work we set
\begin{equation}\label{chi}
\chi({\bf k}\nu)=D({\bf k}\nu).
\end{equation}
The square of the site spin is given by the relation
\begin{equation}\label{S2}
\langle{\bf S}^2\rangle=\frac{3}{2}\frac{T}{N}\sum_{\bf k\nu}D({\bf
k}\nu),
\end{equation}
where Eq.~(\ref{rotinvar}) is taken into account.

\begin{figure}
\centerline{\includegraphics[width=7.5cm]{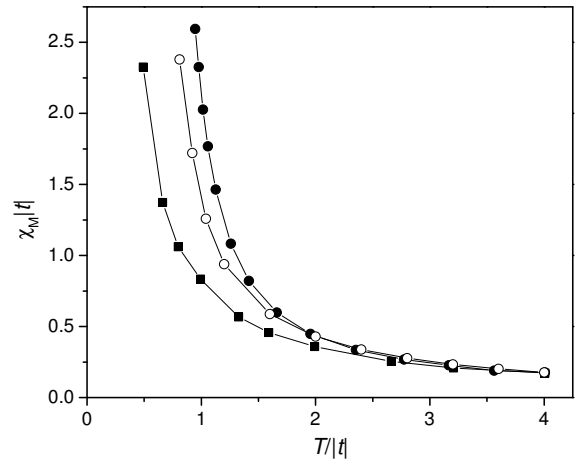}} \caption{The
zero-frequency magnetic susceptibility at ${\bf k}=(\pi,\pi)$ vs.\
temperature at half-filling and $t=-U/4$. Filled squares, filled and
open circles are results of the Monte Carlo simulations,
\protect\cite{Hirsch} random phase approximation, and our calculations,
respectively.}\label{Fig_iv}
\end{figure}

The calculated zero-frequency magnetic susceptibility for ${\bf k}=0$
and half-filling is shown in Fig.~\ref{Fig_iii}. Results obtained in
Monte Carlo simulations\cite{Hirsch} and in the RPA are also shown here
for comparison. The RPA results are described by the
equations\cite{Mahan}
\begin{eqnarray}
&&\chi_{\rm RPA}({\bf k})=\frac{2\chi_0({\bf k})}{1-U\chi_0({\bf k})},
 \nonumber\\
&&\label{RPA}\\
&&\chi_0({\bf k})=-\frac{1}{N}\sum_{\bf k'}\frac{f(t_{\bf k'}-\mu)-
 f(t_{\bf k'+k}-\mu)}{t_{\bf k'}-t_{\bf k'+k}},\nonumber
\end{eqnarray}
where $f(E)=[\exp(E\beta)+1]^{-1}$. Notice that to use the same scale
for the susceptibility as in Ref.~\onlinecite{Hirsch} our calculated
values~(\ref{chi}) in Figs.~\ref{Fig_iii} and \ref{Fig_iv} were
multiplied by the factor 2. Also it should be mentioned that for
$T>2|t|=U/2$ we violate condition~(\ref{condition}); however, the
calculated high-temperature susceptibility is in reasonable agreement
with the Monte Carlo data. It deserves attention that in contrast to
the RPA susceptibility which diverges for low temperatures the
susceptibility in our approach tends to a finite value as it must.

The staggered magnetic susceptibility $\chi_M$ is shown in
Fig.~\ref{Fig_iv}.
As mentioned above, in the used approximation as the temperature
approaches $T_0$ from above, $\chi_M$ tends to infinity which signals
the establishment of the long-range antiferromagnetic order. For
parameters of Fig.~\ref{Fig_iv} $T_0\approx 0.64|t|$. The transition
temperature $T_0$ is finite; however, for the considered range of
parameters $4|t|\leq U\leq 16|t|$ it is always lower than the
respective temperature in the RPA. Accordingly our calculated values of
$\chi_M$ in Fig.~\ref{Fig_iv} are  closer to the Monte Carlo data than
the RPA results.

\begin{figure}
\centerline{\includegraphics[width=7.5cm]{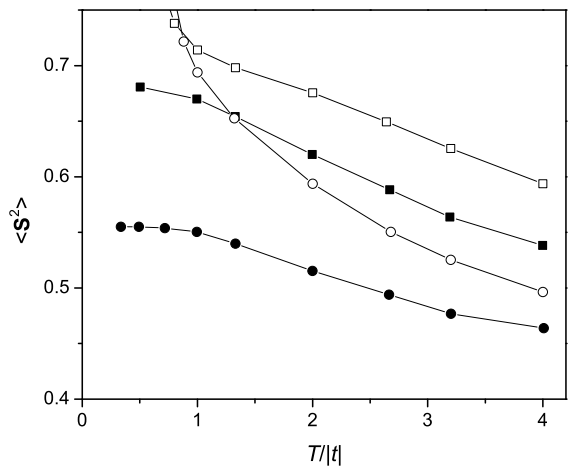}} \caption{The
square of the site spin $\langle{\bf S}^2\rangle$ vs.\ temperature at
half-filling. Filled symbols are data of Monte Carlo simulations,
\protect\cite{Hirsch} open symbols are our results. Squares and circles
correspond to the cases $t=-U/8$ and $t=-U/4$, respectively.}
\label{Fig_v}
\end{figure}

The temperature variation of the square of the site spin,
Eq.~(\ref{S2}), is shown in Fig.~\ref{Fig_v} together with the data of
Monte Carlo simulations.\cite{Hirsch}
As might be expected, the results for the smaller ratio $|t|/U$ more
closely reproduce the data of numerical simulations. For $t=-U/8$ our
calculations replicate the Monte Carlo data for $T\agt |t|$ and the
difference between the two series of results is less than 10 percent.
This difference is at least partly connected with the simplification
made above when irreducible two-leg diagrams were dropped from internal
and external lines of the four-leg diagrams. The difference becomes
even smaller if in accord with the Mermin-Wagner theorem $T_0$ is set
as the zero of the temperature scale and our calculated curve is offset
by this temperature to the left. On approaching $T_0$ our approximation
becomes inapplicable for calculating $\langle{\bf S}^2\rangle$ -- it
starts to grow rapidly and exceeds the maximum value $\frac{3}{4}$.

\begin{figure}
\centerline{\includegraphics[width=7.5cm]{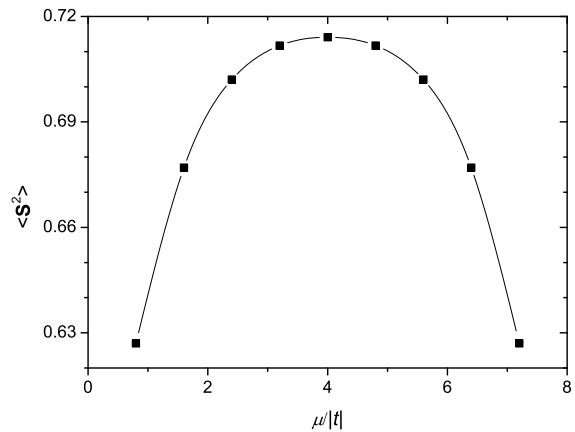}} \caption{The
square of the site spin $\langle{\bf S}^2\rangle$ vs.\ the chemical
potential for $t=-U/8$ and $T=|t|$.} \label{Fig_vi}
\end{figure}

The concentration dependence of $\langle{\bf S}^2\rangle$ near
half-filling is shown in Fig.~\ref{Fig_vi}. The range of the electron
concentration $\bar{n}=\sum_\sigma\langle n_{\bf l\sigma}\rangle$ which
corresponds to the chemical potential shown in this figure spans
approximately $0.8-1.2$ for $t=-U/8$. As would be expected,
$\langle{\bf S}^2\rangle$ decreases rapidly with the departure from
half-filling.

\begin{figure}
\centerline{\includegraphics[width=6.5cm]{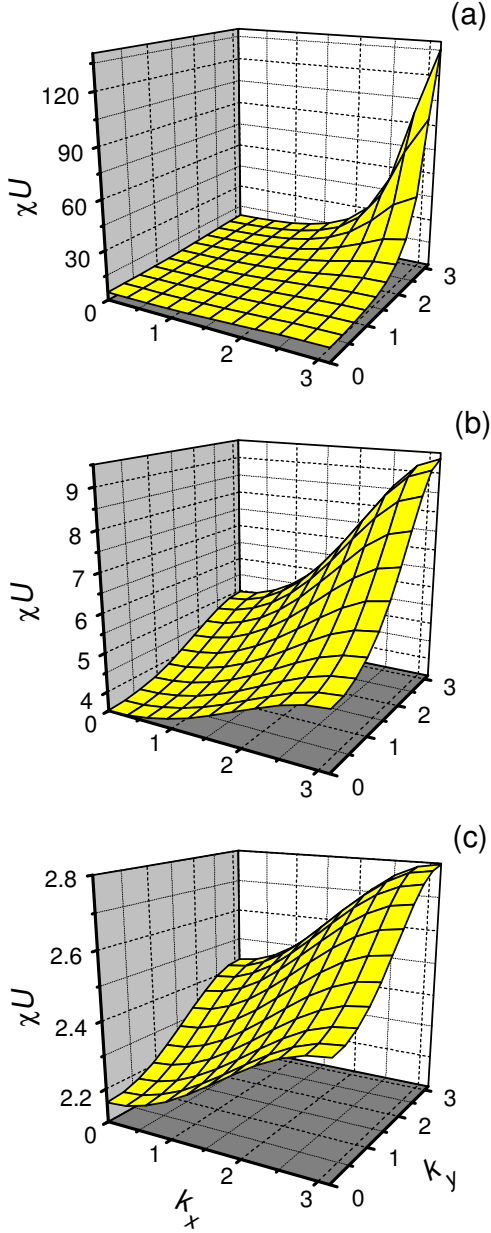}} \caption{(Color
online) The zero-frequency magnetic susceptibility at half-filling for
$t=-U/8$ in a quadrant of the Brillouin zone. (a) $T=0.06U$, (b)
$T=0.1U$, and (c) $T=0.2U$.} \label{Fig_vii}
\end{figure}

The momentum dependence of the zero-frequency susceptibility at
half-filling and its variation with temperature are shown in
Fig.~\ref{Fig_vii}. At half-filling the susceptibility is peaked at the
antiferromagnetic wave vector $(\pi,\pi)$. For temperatures which are
only slightly higher than $T_0$ the peak intensity is large
[Fig.~\ref{Fig_vii} (a)] which leads to a slow decrease of spin
correlations with distance and long correlation lengths (see below).
With increasing temperature the peak intensity of the susceptibility
decreases rapidly [Fig.~\ref{Fig_vii} (b) and (c)] which results in a
substantial reduction of the correlation length. In this case for
distances of several lattice periods the spin correlations are small,
nevertheless they remain antiferromagnetic.

\begin{figure}
\centerline{\includegraphics[width=6.5cm]{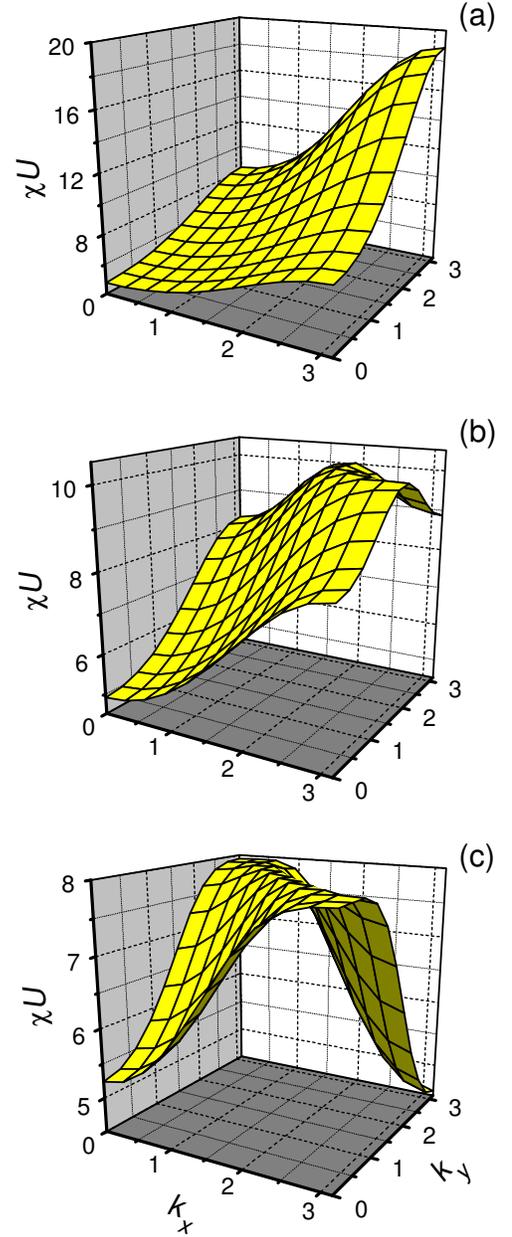}} \caption{(Color
online) The zero-frequency magnetic susceptibility for $t=-U/8$ and
$T=0.06U$ in a quadrant of the Brillouin zone. (a) $\mu=0.2U$, (b)
$\mu=0.15U$, and (c) $\mu=0.1U$.} \label{Fig_viii}
\end{figure}

The situation is changed with the departure from half-filling. The
zero-frequency susceptibility for different electron concentrations is
shown in Fig.~\ref{Fig_viii}. The values of the concentration which
correspond to parts (a) to (c) are $\bar{n}\approx 0.94$, 0.88, and
0.81, respectively. Notice the rapid decrease of the peak intensity of
the susceptibility with doping [cf.\ parts (a) in this and the previous
figure]. Starting from $\bar{n}\approx 0.94$ the susceptibility becomes
incommensurate -- the maximum value of the susceptibility is not
located at $(\pi,\pi)$ -- and the incommensurability parameter, i.e.\
the distance between $(\pi,\pi)$ and the wave vector of the
susceptibility maximum, grows with departure from half-filling. It is
interesting to notice that for $\bar{n}<1$ the zero-frequency
susceptibility diverges when the temperature approaches some critical
temperature in the same manner as it does at half-filling. For $t=-U/8$
and $0.94\alt\bar{n}\leq 1$ the divergence first occurs at $(\pi,\pi)$,
while for smaller electron concentrations it appears at incommensurate
wave vectors. For $\bar{n}<1$ the value of the critical temperature is
less than $T_0$ -- the temperature at which the transition to the
long-range order occurs at half-filling. The critical temperature
decreases with decreasing $\bar{n}$. If in accord with the
Mermin-Wagner theorem we identify $T_0$ with zero temperature we have
to conclude that for $\bar{n}<1$ the system undergoes a virtual
transition at negative temperatures, while for $T\geq 0$ it is governed
by short-range order. In view of the particle-hole symmetry analogous
conclusions can be made for $\bar{n}>1$.

\begin{figure}[!]
\centerline{\includegraphics[width=6.5cm]{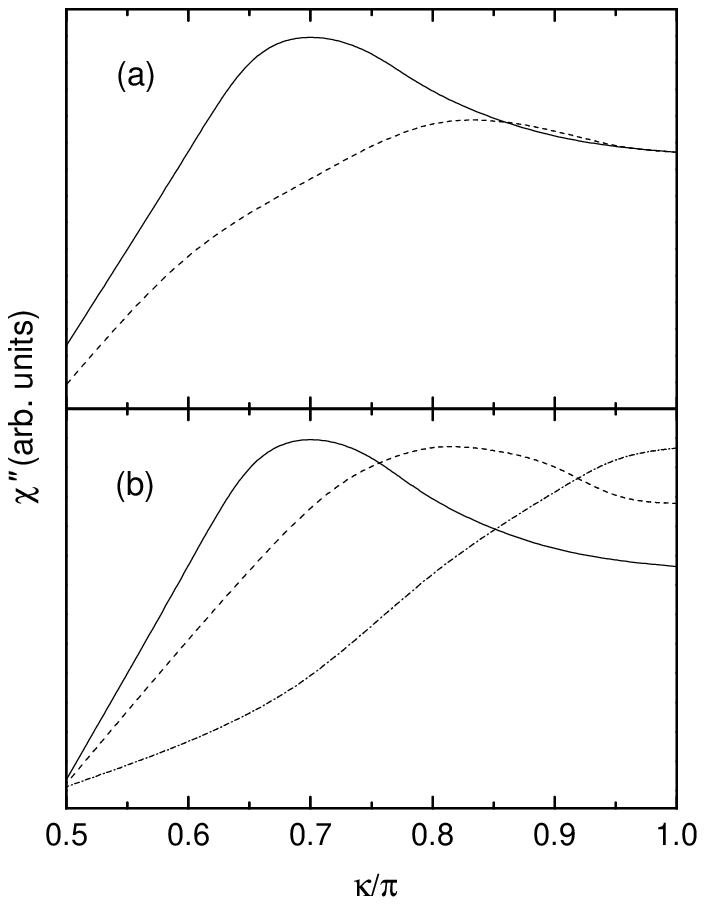}}\vspace{5ex}
\centerline{\includegraphics[width=7cm]{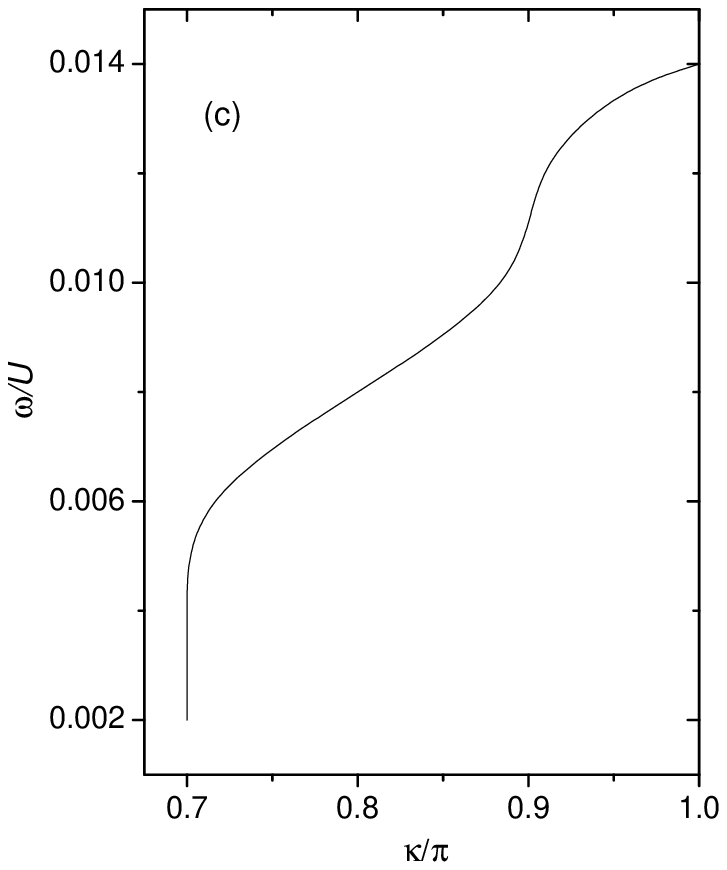}} \caption{(a) The
momentum dependence of $\chi''({\bf k}\omega)$ along the edge [solid
line, ${\bf k}=(\pi,\kappa)$] and diagonal [dashed line, ${\bf
k}=(\kappa,\kappa)$] of the Brillouin zone for $t=-0.11U$,
$\omega=0.002U$ and $\bar{n}\approx 0.88$). (b) The momentum dependence
of $\chi''({\bf k}\omega)$ along the zone edge for $\bar{n}\approx
0.88$ (solid line), $\bar{n}\approx 0.94$ (dashed line), and
$\bar{n}=1$ (dash-dotted line). $t=-0.11U$ and $\omega=0.002U$. (c) The
dispersion of maxima in $\chi''({\bf k}\omega)$ along the zone edge for
$t=-0.11U$ and $\bar{n}\approx 0.88$.} \label{Fig_ix}
\end{figure}

Analyzing equations of the previous section it can be seen that the
momentum dependence of the zero-frequency susceptibility is mainly
determined by the multiplier $y_1({\bf k},\nu=0)$ in the first term on
the right-hand side of Eq.~(\ref{second}). At half-filling the
susceptibility is commensurate, since this term is peaked at
$(\pi,\pi)$ and diverges at this momentum when $T\rightarrow +T_0$, as
the determinant of the system~(\ref{system}) vanishes. At departure
from half-filling the behavior of $y_1$ is governed by the term $b_1$
in this system. The term contains the sum
\begin{equation}\label{sum}
T\sum_n a^2_1(0,n)f_{\bf k}(0,n) =T\sum_n
F^2(n)\bigg[1+\frac{1}{4}tt_{\bf k}F^2(n)\bigg]^{-1},
\end{equation}
where $F(n)=-U[(i\omega_n+\mu)(i\omega_n+\mu-U)]^{-1}$. For
half-filling the sum has a maximum at $(\pi,\pi)$, however with
departure from half-filling the maximum shifts from $(\pi,\pi)$ and the
susceptibility becomes incommensurate.

Together with the zero-frequency susceptibility the imaginary part of
the real-frequency susceptibility,
\begin{equation}\label{imchi}
\chi''({\bf k}\omega)={\rm Im}\, D({\bf k},\omega+i\eta),\quad \eta
\rightarrow +0,
\end{equation}
becomes also incommensurate. This quantity is of special interest,
because it determines the dynamic structure factor measured in neutron
scattering experiments.\cite{Kastner} To carry out the analytic
continuation of $D({\bf k}\nu)$ to the real frequency axis an algorithm
\cite{Vidberg} based on the use of Pad\'e approximants can be applied.
In this calculation 300 values of $D({\bf k}\nu)$ at equally spaced
imaginary frequencies in the upper half-plane were used. The obtained
dependencies of the susceptibility on the momentum for a fixed transfer
frequency $\omega$ and the dispersion of low-frequency maxima in
$\chi''$ are shown in Fig.~\ref{Fig_ix}. The susceptibility is shown in
the first Brillouin zone and can be extended to the second zone by
reflection with respect to the right $y$ axis. As seen from
Figs.~\ref{Fig_ix} (a) and (b), with departure from half-filling
$\chi''({\bf k}\omega)$ becomes incommensurate and the
incommensurability parameter grows with increasing $1-\bar{n}$.

This behavior of the susceptibility $\chi''({\bf k}\omega)$ in the
Hubbard model resembles the low-frequency incommensurate magnetic
response observed by inelastic neutron scattering in lanthanum
cuprates.\cite{Yamada} In these crystals, the incommensurability is
observed both in the normal and superconducting states. For small
transfer frequencies $\omega$ the maxima of the susceptibility are
located on the edge of the Brillouin zone. For the parameters of
Fig.~\ref{Fig_ix} (a) our calculated susceptibility is also peaked on
the zone edge. However, for other parameters the susceptibility on the
diagonal may be comparable to that on the zone edge. This uncertainty
in the position of the susceptibility maxima may be connected with
errors introduced in the calculation results by the procedure of
analytic continuation to real frequencies.

In experiment, for small $\omega$ the incommensurability parameter
grows with the hole concentration $1-\bar{n}$ in the range $0.04\alt
1-\bar{n}\alt 0.12$ and saturates for its larger values. This behavior
of the incommensurability parameter is reproduced in our calculations
[see Fig.~\ref{Fig_ix} (b)] and its values are close to those observed
experimentally. For a fixed hole concentration the incommensurability
parameter decreases with increasing $\omega$ and at some frequency
$\omega_r$ the incommensurability disappears and the susceptibility
$\chi''({\bf k}\omega)$ appears to be peaked at the antiferromagnetic
momentum.\cite{Tranquada} The same behavior is observed in the Hubbard
model [see Fig.~\ref{Fig_ix} (c)]. In lanthanum cuprates for the hole
concentrations $1-\bar{n}\approx 0.12$ the frequency $\omega_r\approx
50$~meV. In Fig.~\ref{Fig_ix} (c) we chose parameters so that
$\omega_r$ was close to this value (for the superexchange constant
$J=4t^2/U\approx 0.15$~eV and $t=-0.11U$ we find $U=3.1$~eV,
$t=0.34$~eV, and $\omega_r=44$~meV). Notice that like in experiment
$\omega_r$ decreases with decreasing $1-\bar{n}$.

A similar incommensurability is observed in YBa$_2$Cu$_3$O$_{7-y}$;
\cite{Arai} however, in this case due to a larger superconducting
temperature and gap the magnetic incommensurability is usually observed
in the superconducting state and the low-frequency part of the
susceptibility is suppressed. As follows from the above discussion, in
the Hubbard model the magnetic incommensurability is a property of
strong electron correlations. The similarity of the mentioned
experimental and calculated results gives ground to consider these
strong correlations as a possible mechanism of the low-frequency
incommensurability observed in experiment. A similar mechanism was
observed for the related $t$-$J$ model in Ref.~\onlinecite{Sherman05}.

\begin{figure}
\centerline{\includegraphics[width=7.5cm]{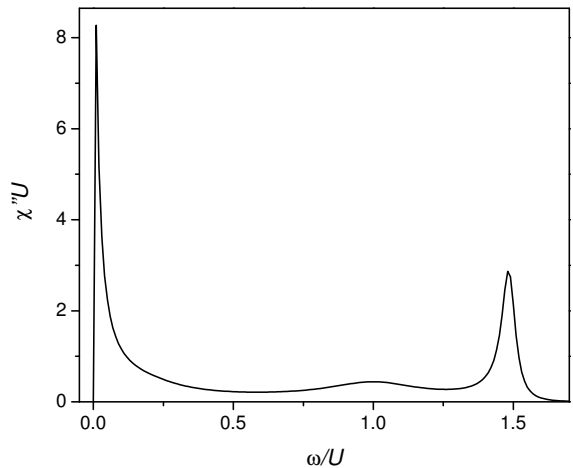}} \caption{The
susceptibility $\chi''({\bf k}\omega)$ for ${\bf k}=(\pi,\pi)$,
$t=-U/8$, $T=0.06U$, and $\mu=0.2U$ ($\bar{n}\approx 0.94$).}
\label{Fig_x}
\end{figure}

In experiment,\cite{Tranquada,Arai} for frequencies $\omega>\omega_r$
the susceptibility $\chi''({\bf k}\omega)$ becomes again incommensurate
such that the dispersion of maxima in $\chi''({\bf k}\omega)$ resembles
a sandglass. The most frequently used interpretations of this
dispersion are based on the picture of itinerant electrons with the
susceptibility calculated in the RPA\cite{Liu} and on the stripe
picture.\cite{Tranquada,Seibold06} In Ref.~\onlinecite{Sherman05} the
sandglass dispersion was obtained in the $t$-$J$ model in the regime of
strong electron correlations without the supposition of the existence
of stripes. In this latter work the part of the sandglass dispersion
for $\omega>\omega_r$ was related to the dispersion of excitations of
localized spins. Similar notion was earlier suggested in
Ref.~\onlinecite{Barzykin}. In our present calculations we did not
obtain this upper part of the dispersion, since the used approximation
does not describe the appearance of localized spins. A typical example
of the frequency dependence of the susceptibility $\chi''({\bf
k}\omega)$ which up to the multiplier $\pi^{-1}$ coincides with the
spin spectral function is shown in Fig.~\ref{Fig_x}.
The susceptibility usually contains several maxima one of which is
located at $\omega\ll U$, while others are placed at frequencies of the
order of $U$. Since the localized spin excitations have frequencies in
the range $0\leq\omega\alt 2J$ where $J=4t^2/U\ll U$, the former
maximum could be taken as a signal for such excitation. However, the
intensity of the maximum usually grows with temperature and with
departure from half-filling. This indicates that the maximum is more
likely due to a bound electron-hole state in which both components
belong to the same Hubbard subband, while in the high-frequency maxima
the components belong to different subbands.

\begin{figure}
\centerline{\includegraphics[width=7.5cm]{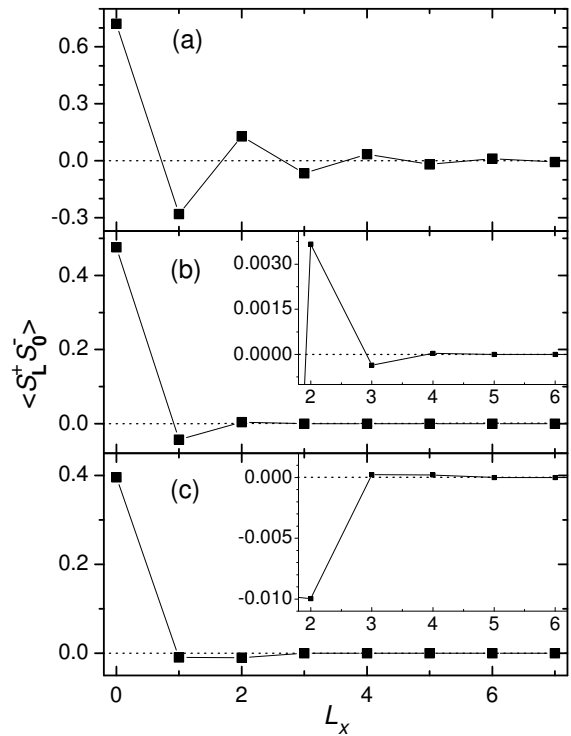}} \caption{The
spin-spin correlator $\langle s^+_{\bf L}s^-_{\bf 0}\rangle$ for ${\bf
L}=(L_x,0)$ and $t=-U/8$. (a) $T=0.06U$, $\mu=0.5U$, (b) $T=0.125U$,
$\mu=0.5U$, and (c) $T=0.06U$, $\mu=0.1U$ ($\bar{n}\approx 0.81$).
Insets in (b) and (c) demonstrate the same data as in the main plots in
a larger scale.} \label{Fig_xi}
\end{figure}

In connection with the Nagaoka theorem\cite{Nagaoka} it is of interest
to investigate the tendency towards the establishment of the
ferromagnetic ordering with departure from half-filling. For a finite
$U$ this problem was investigated by different analytical
methods\cite{Hirsch,Izyumov90,Penn,Kubo} and by Monte Carlo
simulations.\cite{Hirsch} Our results for the spin-spin correlator,
\begin{equation}\label{correlator}
\langle s^+_{\bf L}s^-_{\bf 0}\rangle=\frac{T}{N}\sum_{\bf k\nu}
\cos({\bf kL}) D({\bf k}\nu),
\end{equation}
as a function of the distance $L_x$ between spins are shown in
Fig.~\ref{Fig_xi} for different parameters.
Figure~\ref{Fig_xi} (a) demonstrates the short-range antiferromagnetic
order at half-filling for a temperature which is slightly above $T_0$
(as discussed above in connection with Fig.~\ref{Fig_v}, for such
temperatures the value of $\langle s^+_{\bf 0}s^-_{\bf 0}\rangle$ is
somewhat overestimated by the used approximation). Figure~\ref{Fig_xi}
(b) corresponds also to half-filling to somewhat higher temperature. In
this case the correlations are still antiferromagnetic though they are
characterized by a correlation length which is much shorter than that
in Fig.~\ref{Fig_xi} (a). We have found that the correlation length
diverges when $T\rightarrow T_0$ which indicates the transition to the
long-range antiferromagnetic order. Similar weak antiferromagnetic
correlations were also obtained for moderate departures from
half-filling. Figure~\ref{Fig_xi} (c) corresponds to the lowest filling
$\bar{n}\approx 0.81$ which is allowed by condition~(\ref{condition})
for the given ratio $U/|t|$. According to the mean-field
theory\cite{Hirsch} and the generalized RPA \cite{Izyumov90} in this
case the system has a ferromagnetic ground state. As seen from
Fig.~\ref{Fig_xi} (c), our calculated spin-spin correlations are still
antiferromagnetic even for nearest neighbor spins. This result is in
agreement with Monte Carlo simulations\cite{Hirsch} carried out for the
same parameters. Analogous result was also obtained for $U=16|t|$.
However, a tendency for the establishment of ferromagnetic correlations
can also be seen from the comparison of Figs.~\ref{Fig_xi} (a) and (c)
-- the antiferromagnetic spin correlation on nearest neighbor sites
becomes smaller with doping. For larger ratios of $U/|t|$ we can
ascertain that the correlation changes sign and becomes ferromagnetic.
In particular, it happens at $U/|t|=25$ and $\bar{n}\approx 0.77$. For
these parameters condition~(\ref{condition}) is still fulfilled.

\section{Concluding remarks}
In this article we investigated the magnetic susceptibility of the
two-dimensional repulsive Hubbard model using the diagram technique
developed for the case of strong electron correlations. In this
technique the power series in the hopping constant is used. At
half-filling the calculated temperature dependence of the
zero-frequency susceptibility reproduces adequately key features of
results of Monte Carlo simulations. The uniform susceptibility tends to
a finite value for vanishing temperature. The staggered susceptibility
diverges with decreasing temperature which signals the establishment of
the long-range antiferromagnetic order. The transition temperature is
finite which indicates the violation of the Mermin-Wagner theorem.
However, the transition temperature is always lower than the analogous
temperature in the RPA. Besides, the transition temperature decreases
with the decrease of the ratio $|t|/U$ of the hopping constant and the
on-site repulsion, i.e. the violation of the Mermin-Wagner theorem
becomes less pronounced on enforcing the condition for which the
approximation was developed. For small ratios $|t|/U$ the calculated
square of the site spin differs by less than 10 percent from the data
of Monte Carlo simulations. Also in agreement with Monte Carlo results
we found no evidence of ferromagnetic correlations in the considered
range of electron concentrations $0.8\alt\bar{n}\alt 1.2$ for the
repulsion parameters $8|t|\leq U\leq 16|t|$. However, for larger
$U/|t|$ and $|1-\bar{n}|\approx 0.2$ the nearest neighbor correlations
become ferromagnetic. In the case $U=8|t|$ for $0.94\alt\bar{n}\alt
1.06$ the zero-frequency susceptibility and the imaginary part of the
susceptibility for low real frequencies are peaked at the
antiferromagnetic wave vector $(\pi,\pi)$. For smaller and larger
concentrations these susceptibilities become incommensurate -- momenta
of their maxima are shifted from $(\pi,\pi)$ -- and the
incommensurability parameter, i.e.\ the distance between $(\pi,\pi)$
and the momentum of the maximum susceptibility, grows with departure
from half-filling. With increasing the transfer frequency the
incommensurability parameter decreases and finally vanishes. This
behavior of the susceptibility in the strongly correlated system can
explain the observed low-frequency incommensurate response in the
normal state of lanthanum cuprates.

\begin{acknowledgments}
This work was partially supported by the ETF grant No.~6918 and by the
DFG.
\end{acknowledgments}

\begin{widetext}
\appendix*
\section{}
In this Appendix we prove the symmetry relation~(\ref{rotinvar}). In
the zeroth order of the perturbation expansion for Green's function
(\ref{Dz}) we find
\begin{eqnarray}
D_z^{(0)}({\bf l'\tau',l\tau})&=&\frac{1}{4}\sum_{\sigma\sigma'}\sigma
 \sigma'\Big[K'_2(\tau'\sigma',\tau'\sigma',\tau\sigma,\tau\sigma)
 \delta_{\bf ll'}
 +K_1(\tau'\tau')K_1(\tau\tau)-K_1(\tau'\tau)K_1(\tau\tau')
 \delta_{\bf ll'}\delta_{\sigma\sigma'}\Big]\nonumber\\
&=&\frac{1}{2}\delta_{\bf ll'}\Big[
 K'_2(\tau'\sigma,\tau'\sigma,\tau\sigma,\tau\sigma)
 -K'_2(\tau'\sigma;\tau'\sigma;\tau,-\sigma;\tau,-\sigma)
 -K_1(\tau'\tau)K_1(\tau\tau')\Big], \label{zeroorder}
\end{eqnarray}
where we took into account that the first-order cumulant
$K_1(\tau'\tau)$ does not depend on $\sigma$ and therefore the second
term in the sum vanishes. Up to the multiplier $\frac{1}{2}$ the last
term on the right-hand side of Eq.~(\ref{zeroorder}) coincides with the
respective term in the expansion for Green's function~(\ref{Green}). In
the used diagram technique $K_1$ describes the bare electron Green's
function. Therefore that term contributes to the electron bubble shown
in Fig.~\ref{Fig_i}. From the higher-order terms it can be seen that
inclusion of irreducible two-leg diagrams into the two bare Green's
functions of that term retains the one-to-one correspondence between
terms of the bubble diagrams in $D$ and $D_z$ and the additional
multiplier $\frac{1}{2}$ in the terms of $D_z$.

The second-order cumulant $K_2'$ in Eq.~(\ref{zeroorder}) is defined as
\begin{equation}\label{cumulant2}
K'_2(\tau'\sigma,\tau\sigma,\tau'_1\sigma_1,\tau_1\sigma_1)=\langle
{\cal T}\bar{a}_\sigma(\tau')a_\sigma(\tau)\bar{a}_{\sigma_1}(\tau'_1)
a_{\sigma_1}(\tau_1)\rangle-K_1(\tau',\tau)K_1(\tau'_1,\tau_1)+
K_1(\tau',\tau_1)K_1(\tau'_1,\tau)\delta_{\sigma\sigma_1}.
\end{equation}
This definition is more general than Eq.~(\ref{cumulant}) -- the latter
is obtained from Eq.~(\ref{cumulant2}) if we set $\sigma_1=-\sigma$ and
interchange annihilation operators in $K_2$. After the Fourier
transformation we find
\begin{eqnarray}
&&K'_2(n_1\sigma;n_1+\nu,\sigma;n_2+\nu,\sigma_1;n_2\sigma_1)=
 Z^{-1}\Big\{\beta\Big[\big(\delta_{\nu 0}\delta_{\sigma\sigma_1}
 -\delta_{n_1n_2}\big)
 e^{-E_1\beta}\nonumber\\
&&\quad+Z^{-1}\big(\delta_{\nu 0}-\delta_{n_1n_2}
 \delta_{\sigma\sigma_1}\big)\Big(e^{-(E_0+E_2)\beta}-
 e^{-2E_1\beta}\Big)\Big]F(n_1+\nu)F(n_2)\nonumber\\
&&\quad-\delta_{\sigma,-\sigma_1}e^{-E_0\beta}Ug_{01}(n_1+\nu)
 g_{01}(n_2)g_{02}(n_1+n_2+\nu)
 \big[g_{01}(n_2+\nu)+g_{01}(n_1)\big]\nonumber\\
&&\quad-\delta_{\sigma,-\sigma_1}e^{-E_2\beta}Ug_{12}(n_1+\nu)
 g_{12}(n_2)g_{02}(n_1+n_2+\nu)
 \big[g_{12}(n_2+\nu)+g_{12}(n_1)\big]\nonumber\\
&&\quad+\delta_{\sigma,-\sigma_1}e^{-E_1\beta}\Big[F(n_1+\nu)
 g_{01}(n_2)g_{01}(n_2+\nu)+
 F(n_2)g_{01}(n_1+\nu)g_{01}(n_1)\nonumber\\
&&\quad+F(n_2)g_{12}(n_2+\nu)\big[g_{12}(n_1+\nu)-g_{01}(n_1)\big]+
 F(n_1+\nu)g_{12}(n_1)\big[g_{12}(n_2)-g_{01}(n_2+\nu)\big]\Big]\Big\},
\label{K2p}
\end{eqnarray}
where the notations are the same as in Eq.~(\ref{K2}). From these two
equations it can be seen that
\begin{equation}\label{relforK2}
K_2(n_1,n_1+\nu,n_2+\nu,n_2)=
K'_2(n_1\sigma;n_1+\nu,\sigma;n_2+\nu,\sigma;n_2\sigma)-
K'_2(n_1\sigma;n_1+\nu,\sigma;n_2+\nu,-\sigma;n_2,-\sigma)
\end{equation}
and the analogous equation is fulfilled for the Fourier-transformed
quantities. Thus, zeroth-order terms in the expansions for $D$ and
$D_z$ coincide up to the factor $\frac{1}{2}$.

The next terms in the considered expansions for $D$ and $D_z$ contain
two second-order cumulants and appear in the second order. These terms
read
\begin{eqnarray}
D^{(2)}({\bf l'\tau',l\tau})&=&-\int\!\!\!\int_0^\beta
d\tau_1d\tau_2t_{\bf ll'}t_{\bf l'l}K_2(\tau',\tau',\tau_1,\tau_2)
 K_2(\tau_2,\tau_1,\tau,\tau), \label{Dt2}\\
D_z^{(2)}({\bf l'\tau',l\tau})&=&-\frac{1}{4}
 \sum_{\sigma\sigma'\sigma_1}
 \int\!\!\!\int_0^\beta d\tau_1d\tau_2t_{\bf ll'}t_{\bf l'l}
 K'_2(\tau'\sigma',\tau'\sigma',\tau_1\sigma_1,\tau_2\sigma_1)
 K'_2(\tau_2\sigma_1,\tau_1\sigma_1,\tau\sigma,\tau\sigma).\label{Dl2}
\end{eqnarray}
Using twice relation~(\ref{relforK2}) in Eq.~(\ref{Dl2}) one can see
that $D^{(2)}=2D^{(2)}_z$. Analogous equations for higher order terms
can be proved in the same manner. Thus, relation~(\ref{rotinvar}) is
fulfilled.
\end{widetext}

\end{document}